\documentstyle[aps,preprint]{revtex}

\begin{document}
\draft

\title{Resonance transmission through a single atom}

\author{Jos\'e-Luis Mozos$^{(1,a)}$, C.C. Wan$^{(1)}$,
Gianni Taraschi$^{(1)}$, Jian Wang$^{(2)}$, and Hong Guo$^{(1)}$}
\address{(1) Centre for the Physics of Materials and
Department of Physics, McGill University,\\
Montreal, Quebec, Canada H3A 2T8.\\
(2) Department of Physics, The University of Hong Kong, Pokfulam Road, 
Hong Kong.}

\maketitle

\begin{abstract}

When a single atom is sandwiched in between two electrodes, an 
{\it atomic} tunneling device may be realized depending on the distance
between the atom and the electrodes. We have performed 
first-principles pesudopotential calculation in conjunction with a
three-dimensional evaluation of the quantum scattering matrix for
such a device. We predict the quantum conductance and resonance 
transmission properties of the Si and Al atomic tunneling devices.

\end{abstract}


\medskip

Exploiting the band gap differences of various compound semiconductors
such as GaAs and AlGaAs, resonance tunneling devices have been
fabricated and extensively investigated since the original work of 
Esaki\cite{esaki}. A tunneling device may give rise to negative 
differential resistance due to a quantum resonance, leading to useful 
electronic applications. In this work, we propose and 
theoretically analyze a different kind tunneling device, namely
{\it atomic} scale tunneling systems. Rather than resonant transmitting 
through the quasi-bound states inside a double barrier tunneling
structure, in the atomic system a resonance is through atomic orbitals.
Since atomic orbitals have energy spacings in the range of eV, the quantum
resonance will not be smeared out by a room temperature. Hence in principle
room temperature quantum resonance tunneling devices can be achieved.  
Our work is encouraged by the measurements using scanning 
tunneling microscope (STM) on quantum conduction through atomic scale
tip-substrate systems\cite{pascual}, where
quantized conductance is indeed observed at room temperature.

We propose an atomic tunneling structure as shown schematically in the inset
of Fig. (3). The essential part of the structure consists of two metallic
electrodes sandwiching a single atom in between. The whole system could
in principle be fabricated on top of an insulating substrate by 
micro-fabricating two electrodes and placing an atom in between using the 
atomic manipulation ability of STM. Fixing the distance between the atom 
and the electrode, $d$, vacuum barriers can be established on the two sides 
of the atom, thus a double-barrier atomic tunneling (DBAT) device is 
established\cite{lang1}. The idea of using vacuum barriers to establish a 
tunnel junction is widely used in STM\cite{lyo}, but an DBAT 
structure as an atomic quantum functional device is very interesting 
indeed. The main task of our work, to be presented below, is 
to predict quantum conductance of the DBAT device for various values of 
$d$, to investigate the formation of a atomic quantum point contact as 
$d$ is reduced thus diminishing the vacuum barriers, and to examine 
issues which arise when transport is mediated by atomic levels.
Although we have concentrated on DBAT devices operated on a single atom, 
the idea can be extended to clusters, molecules, or other groups of atoms.

To capture the atomic degrees of freedom, we have combined the
{\it ab initio} pseudopotential total energy method with a
three-dimensional (3D) quantum scattering evaluation of the transmission
probabilities\cite{wan1}. First, we solve the {\it ground state}
properties of the atom {\it and} the two leads by minimizing the
Kohn-Sham total energy functional using a plane-wave basis 
set\cite{payne}. The equilibrium analysis produces the self-consistent 
effective potential $V_{eff}({\bf r})\equiv \delta U/\delta \rho({\bf r})$ 
which is seen by all the electrons. Here $U[\rho]$ is the total 
self-consistent potential energy while $\rho$ the electron density. 
Second, we evaluate the scattering matrix of an electron traversing
the system defined by $V_{eff}$, by solving a 3D scattering 
problem using a transfer matrix technique\cite{sheng}. From the scattering
matrix we obtain conductance from the Landauer formula\cite{landauer}.

We have focused on DBAT devices made of Si and Al atoms and used
pseudopotentials of Refs. \onlinecite{ihm,goodwin} for the core, 
and the parameterization of Ref. \onlinecite{teter} for the 
exchange-correlation term. The leads are modeled by the jellium model. 
For Si, a jellium lead has a cross section area of 
$7.25\times 7.25(a.u.)^2$, length $L=30.78$a.u., and its charge is
specified\cite{lang2} by $r_s\approx 2.0$a.u., mimicking metallic leads. 
The supercell volume used in our plane-wave based {\it ab initio}
calculations is $21.77\times 21.77\times 2(L+d)$ (a.u.)$^3$.
For the Al DBAT device, the lead has 
a cross section area of $8.79\times 8.79(a.u.)^2$, length $L=23.57a.u.$, 
charge density specified by $r_s\approx 2.07a.u.$. The supercell size 
for the Al DBAT device is $16.67\times 16.67\times 2(L+d)$ (a.u.)$^3$. 
The DBAT setup with square shaped leads has a symmetry of space group D4h. 
We have used an upper energy cutoff of $8$ Rydbergs\cite{foot1}.

For a single adatom on top of a high density jellium substrate, previous 
calculations showed\cite{lang3} that the equilibrium jellium-atom bond 
length is $2.3$a.u. for Si, and $2.6$a.u. for Al\cite{lang2}. Hence for 
values of $d$ greater than these values, a vacuum barrier may be 
established. Fig. (1) shows the $V_{eff}$ for a Si DBAT device with 
$d=6.9$ a.u.. $V_{eff}$ in the 3D leads is essentially a potential well 
with a depth $\sim -0.50$a.u. below the Fermi level of the system. From 
$V_{eff}$ it is clear that the atom is quite isolated from the leads 
by the vacuum barriers. Reducing $d$ to $3.45$ and $4.6$a.u., a 
lower vacuum barrier is obtained as expected. Similar behavior is found
for Al DBAT devices. In Fig. (1), the sharp peak at the atomic position 
reflects the repulsive atomic core, and surrounding the core there is 
the usual attractive part of the atomic potential.

Figs. (2,3) show the conductance $G$ for a number of Si and Al DBAT 
devices which differ by the jellium-atom junction distance $d$. When 
$d$ takes the value of equilibrium jellium-atom bond length, $G(E)$ 
shows the expected ``quantized'' conductance\cite{foot2} for quasi-1D 
quantum wires. However a striking result is the apparent resonance 
transmission when the atom is isolated inside the DBAT device by the 
barriers. For both Si and Al systems, the larger the $d$, the sharper the 
resonances. This is because a larger $d$ corresponds to a more isolated atom.
The resonance peaks must come from the atomic orbitals since the DBAT
device is operating on the electronic structure of the single atom. For
a Si atom, the valence configuration is $3s^2 3p^2$. Thus the
equilibrium ``Fermi'' energy for an isolated atom should be located at 
the $3p$ atomic level. For all the Si DBAT devices, our calculated
$E_f\approx -0.10$a.u., which is very close to the higher energy
peak position of Fig. (2). This allows us to identify the resonance
peak near $-0.09\sim-0.10$a.u. as due to the $3p$ atomic orbital. 
For the largest $d$ we studied, {\it e.g.} $d=6.9$a.u., the sharp $3p$ 
resonance is split into two nearby peaks as shown in Fig. (2). 
It is not difficult to understand this splitting when we realize that the 
$3p$ state is triplely degenerate, and its rotational symmetry is broken 
by the presence of the squared shaped leads which is along the $z$-direction. 
Hence for the single $3p_z$ state, the resonance is marked by 
$G(E_{3p_z})\approx 1$ in units of $2e^2/h$.  On the other hand, the D4h
space group respects the rotational symmetry in the $x-y$ direction,
thus the degenerate $3p_{x,y}$ states give $G(E_{3p_{xy}})\approx 2$. 
The reason that these peaks are slightly higher than $1$ and $2$ is due
to the partial overlap of the states, as is clearly seen in Fig. (2).
When the lead-atom distance $d$ is smaller, such as $d=3.45$a.u. and 
$4.6$a.u., the atom is not as isolated. For these cases, the resonance 
width is larger which smears out the $3p$ splits. As a result, the three 
states of $3p$ atomic orbital gives a $G(E_{3p})=3\times (2e^2/h)$ resonance.

After understanding the $3p$ resonance peak, it is clear that 
the lower energy peak of Fig. (2) must result from the
$3s$ atomic state. Indeed, since $3s$ is a singlet, we have
$G(E_{3s})=1\times (2e^2/h)$. For $d=6.9$a.u., the $3s$ peak is
extremely sharp, we reproduce it in the inset of Fig. (2). A crucial 
numerical test of these results is the distance between the peak positions. 
For a single Si atom, LDA calculation\cite{database} gives energy
level spacing between the $3s$ and $3p$ states to be $0.245$a.u.. 
Our conductance calculation, as shown in Fig. (2), gives the $3s-3p$ 
resonance distance in almost perfect agreement with the LDA spectra,
with only small differences due to the level splitting and the presence
of the leads. Hence it is unambiguous that the transport in a
single atom DBAT device is mediated by the atomic levels.  For Al DBAT 
devices, as shown in Fig. (3), all the results give a physical
picture which is consistent with that of the Si system, including the 
quantitative value of the 3s-3p resonance peak distance. For different 
values of $d$, there is a small but noticeable shift of the relative peak 
positions (see Figs. (2,3)). This is due to the coupling of the atom to 
the leads, and can be understood from previous model calculations\cite{wang}. 
Finally, Figs. (2,3) shows the formation of a 
transmissive quantum point contact: by reducing $d$ to the equilibrium bond 
length between the atom and the leads, the resonance transmission crosses 
over to the usual quantized conductance with a quasi-1D nature.

Certain features of the {\it ab initio} results can be understood 
analytically. First, it can be generally shown\cite{laughlin} that for 
3D resonance tunneling through a symmetric double barrier potential, 
the peak value of the conductance $G(E)$ for an isolated 
singlet state is universally $2e^2/h$. The 3s peak shows this result. 
Second, using the discussion of Ref. \onlinecite{laughlin} and Fermi's 
golden rule, it's not difficult to obtain the resonance width as
\[\Gamma= 2 \pi \sum_j |M_j|^2 \delta(E_r-E_j) = \hbar \sum_j w_j, \]
where $w_j$ is the tunneling rate from the atomic resonance level
$E_r$ to the lead level $E_j$, and $M_j$ is the corresponding
tunneling matrix element from $E_r$ to $E_j$. Since the tunneling 
rate decreases exponentially with the barrier 
size, we expect the resonance widths to do the same, which 
is clearly seen in our numerical results. Third, we can understand why 
the $3p_z$ resonance peak width is wider than that of the $3p_{xy}$ peak. 
Using a first-order approximation for the tunneling matrix\cite{chen}
\[M_j = \int \chi_j^\ast V_{lead} \phi_r ~ d^3r,\]
where $\chi_j$ is the lead state, $\phi_r$ is the atomic state, and 
$V_{lead}$ is the potential due to the lead. By approximating 
the lead wavefunctions with finite square well wavefunctions, and $V_{lead}$
as a potential well, we find $\Gamma_{p_z}/\Gamma_{p_{xy}} =6.2$ for 
$d=6.9 a.u$. This agrees well with the value of $6.0$ from our 
{\it ab initio} calculation. We thus conclude, that the $3p_z$ resonance 
peak width is wider than that of the $3p_{x,y}$ peak, because the 
overlap of the $3p_z$ state with the lead wavefunctions is much larger 
than the corresponding overlap for the $3p_{xy}$ states.

In summary, we have proposed and theoretically investigated atomic scale
double barrier tunneling devices. The quantum resonance transmission in
this system is through atomic orbitals, thus conductances obtained
characterizes the conduction through a single atom and reveals the 
valence atomic spectra. Finally we comment that the curves of
$G=G(E)$ are useful for giving a first estimate of the electrical
current as a function of a voltage across the DBAT device: one obtains the
current in the usual fashion by integrating $G(E)$ over energy convoluted 
with a Fermi function. The importance of the atomic DBAT device lies 
in the fact that the quantum resonances are observable at room temperature 
due to the large energy scales (eV) associated with the atomic system. 
Indeed, with the nonlinear conductance behavior of the DBAT device, many
applications can be envisioned.

\section*{Acknowledgments}
We gratefully acknowledge financial support by NSERC of Canada and FCAR of
Quebec. J. Wang is supported by a RGC grant of the Hong Kong Government
under grant number HKU 261/95P, and a research grant from the 
Croucher Foundation. We thank the Computing Center of University of
Hong Kong for a substantial CPU allocation on their IBM SP2 parallel
computer for the numerical analysis presented here.


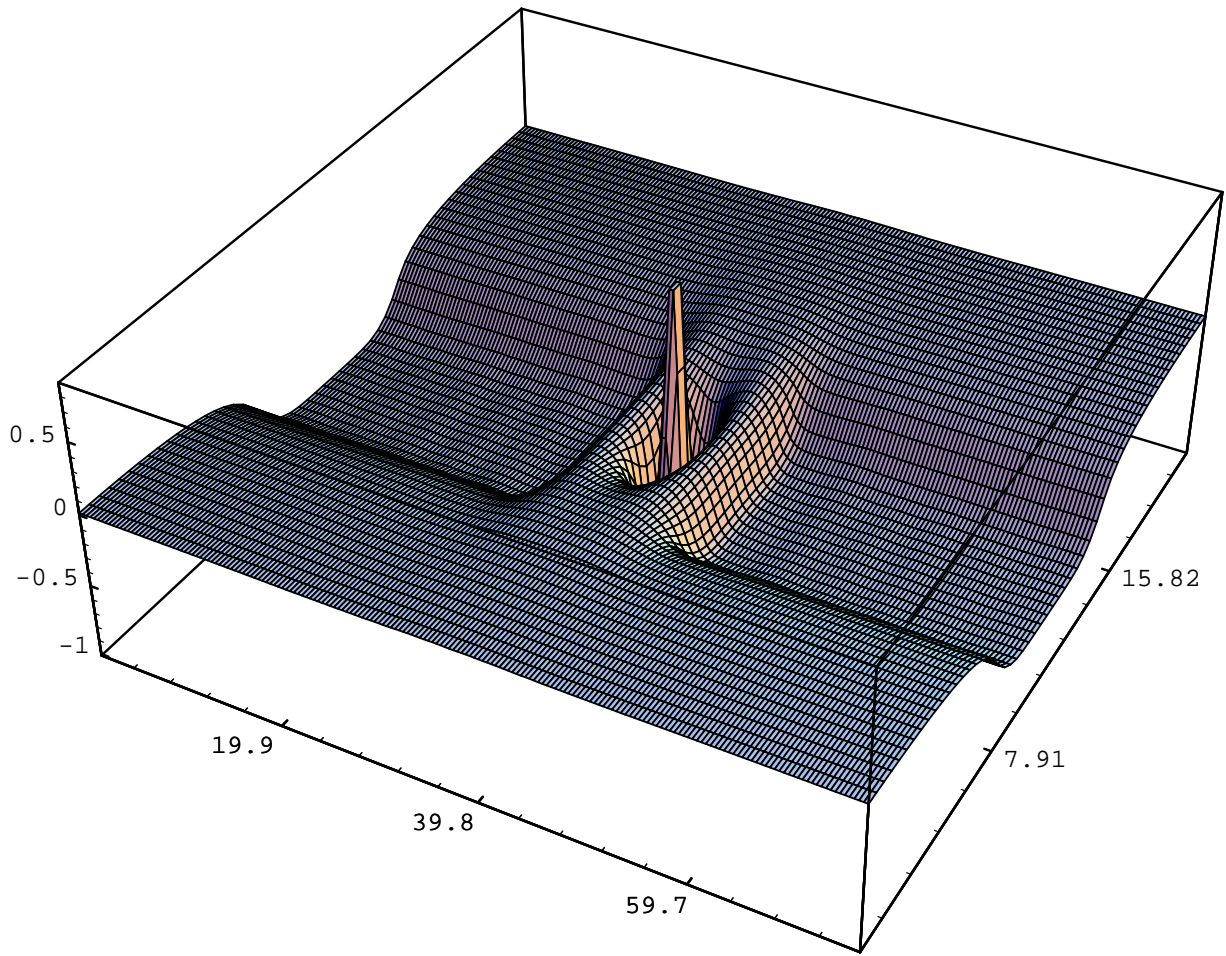
\begin{figure}
\caption{Effective potential $V_{eff}$ for a single atom Si DBAT device.  
The atom-lead distance is fixed at $d=6.9$a.u.. The vacuum barriers are
clearly seen.
}
\label{fig1}
\end{figure}

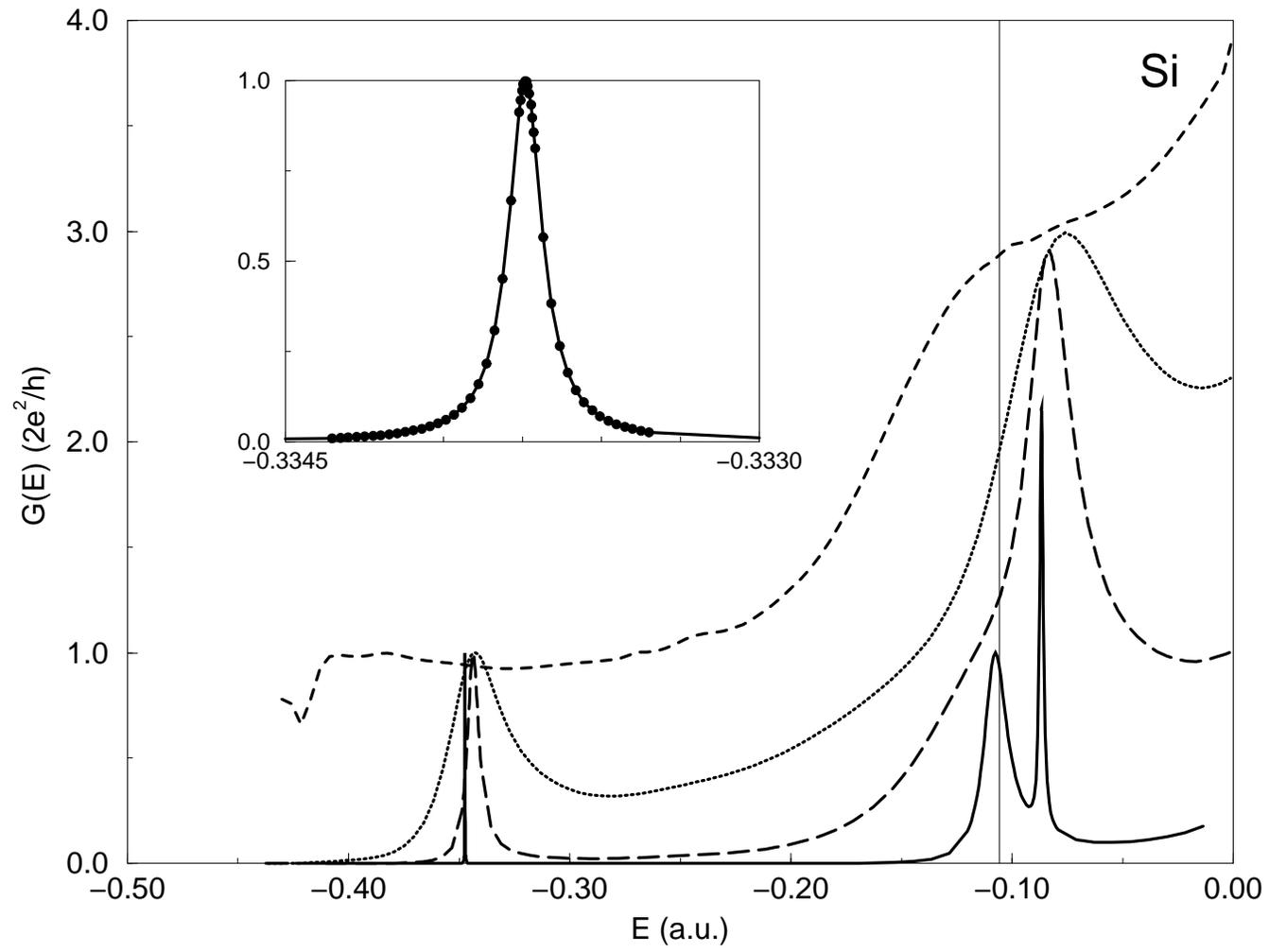
\begin{figure}
\caption{Conductance $G(E)$ as a function of incoming electron energy $E$ 
for Si DBAT devices. Dashed line: for $d=2.3$a.u., which is the
equilibrium atom-leads distance;  dotted line: for $d=3.45$a.u.; long
dashed line: for $d=4.6$a.u.; solid line: for $d=6.9$a.u..
Inset: resolving the $3s$ resonance peak for the $d=6.9$a.u. case.
The thin vertical line indicates the calculated Fermi level of the system.
}
\label{fig2}
\end{figure}

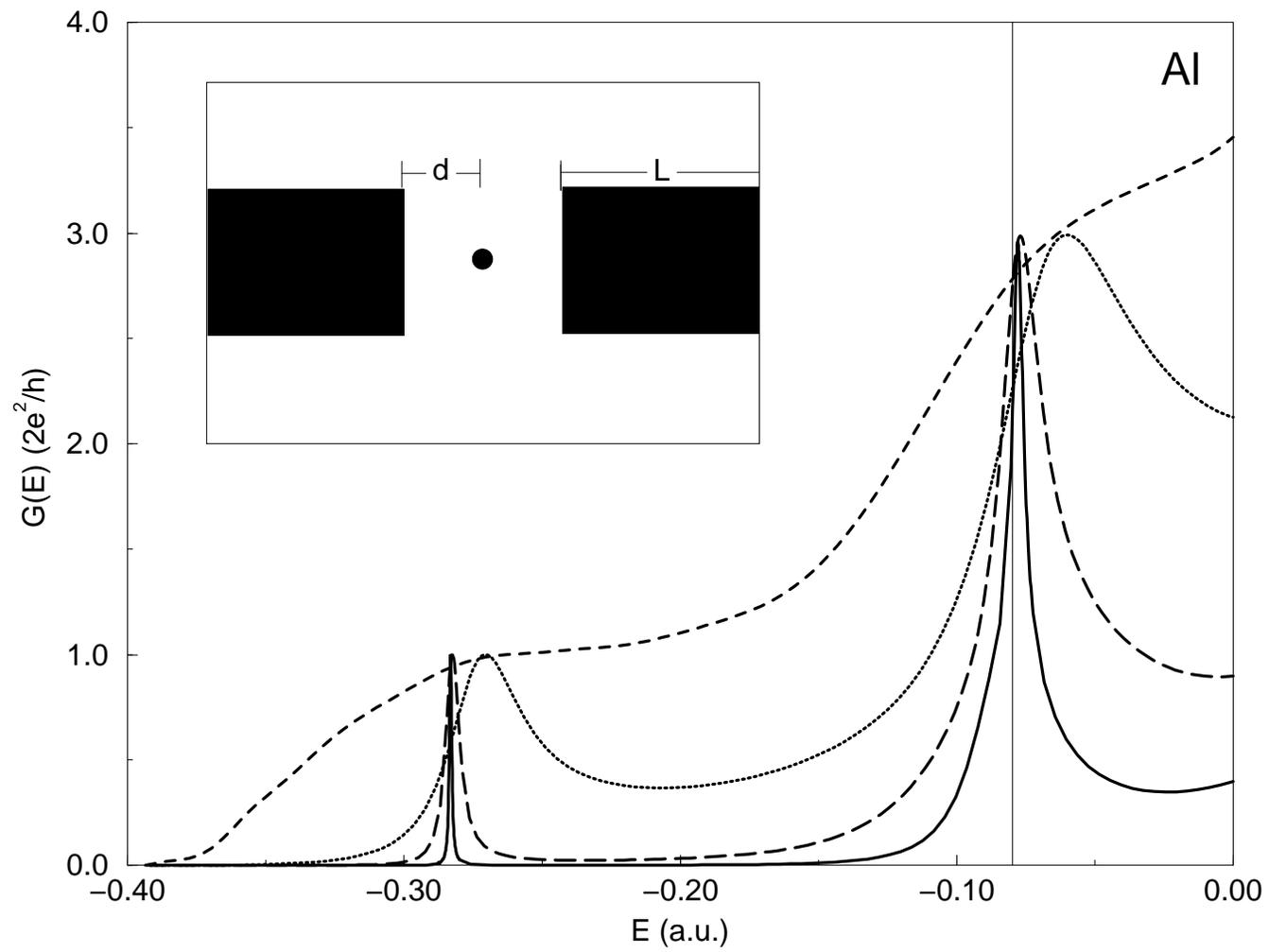
\begin{figure}
\caption{Conductance $G$ for Al DBAT devices. Dashed line: 
for $d=2.6$a.u.; dotted line: $d=3.9$a.u.; long dashed line:
$d=5.2$a.u.; solid line: $d=6.5$a.u.. Inset: schematic plot of the 
atomic DBAT device: an atom is sandwiched in between two metallic wires. 
The thin vertical line indicates the calculated Fermi level of the system.
}
\label{fig3}
\end{figure}

\begin{thebibliography}{00}
\bibitem[]{}
$^{a)}$Present address: 
Fritz-Haber-Institut der Max-Planck-Gesellschaft, Farada yweg 4-6,
D-14195 Berlin-Dahlem, Germany

\bibitem{esaki}
L. Esaki, Phys. Rev. {\bf 109}, 603 (1958).

\bibitem{pascual}
J.I. Pascual, {\it et. l.}
Science, {\bf 267}, 1793 (1995);
E.S. Snow, D. Park, and P.M. Campbell, Appl. Phys. Lett. {\bf 69}, 269
(1996); Ali Yazdani, D.M. Eigler, N.D. Lang, Science, Vol. {\bf 272}, 
1921 (1996).

\bibitem{lang1}
Recently Lang has discussed another possible way to establish a 
tunnel junction by inserting spacer atoms between the electrodes to 
isolate the metal atom which mediates the transport.  See,
N.D. Lang, Phys. Rev. B. {\bf 55}, 9364 (1997).

\bibitem{lyo}
I. W. Lyo and Ph. Avouris, Science {\bf 245}, 1369 (1989);
Ph. Avouris, I.W. Lyo, F. Bozso and E. Kaxiras, J. Vac. Sci. Technol. A
{\bf 8}, 3405 (1990).

\bibitem{wan1}
C.C. Wan, J\'ose-Luis Mozos, Gianni Taraschi, Jian Wang and Hong Guo,
Appl. Phys. Lett. {\bf 71}, 419 (1997).

\bibitem{payne}
M.C. Payne, M.P. Teter, D.C. Allan, T.A. Arias and J.D. Joannopoulos,
Rev. Mod. Phys.  {\bf 64}, 1045 (1992).  

\bibitem{sheng}
W.D. Sheng, J.B. Xia, Phys. Letts. A {\bf 220}, 268(1996).

\bibitem{landauer}
R. Landauer, IBM J. Res. Dev. {\bf 1}, 233 (1957). 

\bibitem{ihm}
J. Ihm and M.L. Cohen, Solid State Commun., {\bf 29}, 711(1979).

\bibitem{goodwin}
L. Goodwin, R.J. Needs and V. Heine, J. Phys.: Condens. Matter {\bf 2},
351 (1990).

\bibitem{teter}
S. Goedecker, M. Teter and J. Hutter, Phys. Rev. B. {\bf 54}, 1703 (1996).

\bibitem{lang2}
N.D. Lang, Phys. Rev. B. {\bf 52}, 5335 (1995);

\bibitem{foot1}
We present results in atomic units through out this work.

\bibitem{lang3}
N.D. Lang and A.R. Williams, Phys. Rev. B {\bf 18}, 616 (1978).

\bibitem{foot2}
The quantization plateau is not perfect, because the single atom
junction is too short to establish perfect quasi-1D transport channels.
For a discussion concerning the establishment of quasi-1D behavior, 
see Ref. \onlinecite{wan1}.

\bibitem{database}
We obtained the DFT results from the NIST Basic Reference Data for 
Electronic Structure Calculations Database 
(http://math.nist.gov/DFTdata/). We have also independently verified
the quoted value.

\bibitem{wang}
Yongjiang Wang, Jian Wang and Hong Guo, Appl. Phys. Lett. {\bf 65},
1793 (1994).

\bibitem{laughlin}
V. Kalmeyer and R.B. Laughlin, Phys. Rev. B {\bf 35}, 9805 (1987).

\bibitem{chen}
C.J. Chen, {\it Introduction to Scanning Tunneling Microscopy}
(Oxford, New York, 1993), p. 68.

\end{thebibliography}
\end{document}